\documentstyle[aps,preprint,tighten,eqsecnum,epsfig,floats,prd]{revtex}

\newcommand{\del}{\partial}
\newcommand{\oa}[1]{${\cal O}(a^{#1})$}

\newcommand{\rf}[4]{{\em {#1}} {\bf #2}, #3 (#4)}

\newcommand{\pr}{Phys.\ Rev.\ }

\newcommand{\physl}{Phys.\ Lett.\ }

\newcommand{\np}{Nucl.\ Phys.\ }
\newcommand{\beq}{\begin{equation}}
\newcommand{\eeq}{\end{equation}}
\newcommand{\beqa}{\begin{eqnarray}}
\newcommand{\eeqa}{\end{eqnarray}}

\newcommand{\Tr}{{\rm Tr}\,}
\newcommand{\muhat}{\hat{\mu}}
\newcommand{\qhat}{\hat{q}}

\newbox\rotbox

\begin{document}

\preprint{\vbox{
\rightline{ADP-00-10/T396}}}

\draft

\title{Infrared Behavior of the Gluon Propagator on a\\ 
       Large Volume Lattice}

\author{Fr\'ed\'eric D.R.\ Bonnet\footnote{
E-mail:~fbonnet@physics.adelaide.edu.au},
Patrick O.\ Bowman\footnote{
E-mail:~pbowman@physics.adelaide.edu.au \hfill\break
\null\quad\quad
WWW:~http://www.physics.adelaide.edu.au/$\sim$pbowman/},
Derek B.\ Leinweber\footnote{
E-mail:~dleinweb@physics.adelaide.edu.au ~$\bullet$~ Tel:
+61~8~8303~3548 ~$\bullet$~ Fax: +61~8~8303~3551 \hfill\break
\null\quad\quad
WWW:~http://www.physics.adelaide.edu.au/theory/staff/leinweber/} \\
and Anthony G.\ Williams\footnote{
E-mail:~awilliam@physics.adelaide.edu.au ~$\bullet$~ Tel:
+61~8~8303~3546 ~$\bullet$~ Fax: +61~8~8303~3551 \hfill\break
\null\quad\quad
WWW:~http://www.physics.adelaide.edu.au/theory/staff/williams.html}}
\address{Special Research Centre for the Subatomic Structure of Matter
and The Department of Physics and Mathematical Physics, University of
Adelaide, Adelaide SA 5005, Australia}
\date{\today}

\maketitle

\begin{abstract} 
The first calculation of the gluon propagator using an \oa{2} improved
action with the corresponding \oa{2} improved Landau gauge fixing
condition is presented.  The gluon propagator obtained from the
improved action and improved Landau gauge condition is compared with
earlier unimproved results on similar physical lattice volumes of
$3.2^3 \times 6.4 \text{ fm}^4$.  We find agreement between the
improved propagator calculated on a coarse lattice with lattice
spacing $a = 0.35$ fm and the unimproved propagator calculated on a fine
lattice with spacing $a = 0.10$ fm.  This motivates us to calculate
the gluon propagator on a coarse large-volume lattice $5.6^3 \times
11.2 \text{ fm}^4$.  The infrared behavior of previous studies is
confirmed in this work.  The gluon propagator is enhanced at
intermediate momenta and suppressed at infrared momenta.  Therefore
the observed infrared suppression of the Landau gauge gluon propagator is not 
a finite volume effect.
\end{abstract}




\section{Introduction}
\label{sec:intro}

There has been considerable interest in the infrared behavior of the 
gluon propagator, as a probe into the mechanism of confinement and
as input for other calculations.  Lattice gauge theory is an excellent means
to study such nonperturbative behavior.  See, for example, Ref.~\cite{Man99} 
for a recent survey. 

The infrared part of any lattice calculation may be affected by the
finite volume of the lattice.  Larger volumes mean either more lattice
points (with increased computational cost) or coarser lattices (with
corresponding discretization errors).  The desire for larger physical
volumes thus provides strong motivation for using improved actions.
Improved actions have been shown to reduce discretization effects
\cite{Alf95}, although some concerns have been expressed that coarse
lattices may miss important instanton physics.

In this study, no change is seen in the infrared gluon propagator, even with
a lattice spacing as coarse as 0.35 fm.
We find the gluon propagator to be less singular than $\frac{1}{q^2}$
in the infrared.  Our results suggest that the gluon propagator is
infrared finite, although more data is needed in the far infrared to
be conclusive.  This behavior is similar to that observed in
three-dimensional SU(2) studies \cite{Cucc99}.

\section{\boldmath\oa{2} Improvement}

The ${\cal O}(a^2)$ tadpole-improved action is defined as
\beq
S_G=\frac{5\beta}{3}\sum_{\text{pl}}{\cal R}eTr(1-U_{\text{pl}}(x)) 
	- \frac{\beta}{12u_{0}^2}\sum_{\text{rect}}
	{\cal R}eTr(1-U_{\text{rect}}(x)),
\label{gaugeaction}
\eeq
where the operators $U_{\text{pl}}(x)$ and $U_{\text{rect}}(x)$ are defined 
\beqa
U_{\text{pl}}(x) & = & U_{\mu}(x)U_{\nu}(x+\hat{\mu})
		U^{\dagger}_{\mu}(x+\hat{\nu}) U^\dagger_{\nu}(x),
\eeqa
and
\beqa
U_{\text{rect}}(x) & = & U_{\mu}(x)U_{\nu}(x+\hat{\mu})
		U_{\nu}(x+\hat{\nu}+\hat{\mu})
		U^{\dagger}_{\mu}(x+2\hat{\nu})U^{\dagger}_{\nu}(x+\hat{\nu})
		U^\dagger_{\nu}(x) \nonumber \\
	& + & U_{\mu}(x)U_{\mu}(x+\hat{\mu})U_{\nu}(x+2\hat{\mu})
		U^{\dagger}_{\mu}(x+\hat{\mu}+\hat{\nu})
		U^{\dagger}_{\mu}(x+\hat{\nu})U^\dagger_{\nu}(x).
\eeqa
The link product $U_{\text{rect}}(x)$ denotes the rectangular
$1\times2$ and $2\times1$ plaquettes.  For the tadpole (mean-field)
improvement parameter \cite{tadpole} we use the plaquette measure
\beq
u_0=\left(\frac{1}{3}{\cal R}eTr<U_{\text{pl}}>\right)^{\frac{1}{4}}.
\eeq
Eq.~(\ref{gaugeaction}) reproduces the continuum action as
$a\rightarrow{0}$, provided that $\beta$ takes the standard value of
$6/g^2$.  Note that our $\beta=6/g^2$ differs from that used
in~\cite{Alf95,Lee,Woloshyn}.  Multiplication of our $\beta$ in
Eq.~(\ref{gaugeaction}) by a factor of $5/3$ reproduces their
definition.  ${\cal O}(g^2a^2)$ corrections to this action are
estimated to be of the order of two to three percent~\cite{Alf95}.

Gauge configurations are generated using the
Cabbibo-Marinari~\cite{Cab82} pseudoheat-bath algorithm with three
diagonal $SU(2)$ subgroups.  The mean link, $u_0$, is averaged every
10 sweeps and updated during thermalization.  Representative gauge
fields are selected after 5000 thermalization sweeps.

Gauge fixing on the lattice is achieved by maximizing a functional, the 
extremum of which implies the gauge fixing condition.  The usual Landau 
gauge fixing functional \cite{cthd} is 
\begin{equation}
{\cal F}^{G}_{1}[\{U\}] = \sum_{\mu, x}\frac{1}{2u_0} \mbox{Tr} \, \left\{
U^{G}_{\mu}(x) + U^{G}_{\mu}(x)^{\dagger} \right\},
\label{eqn:f1}
\end{equation}
where
\begin{equation}
U^{G}_{\mu}(x) = G(x) U_{\mu}(x) G(x+\hat{\mu})^{\dagger},
\end{equation}
and
\begin{equation}
G(x) = \exp \left\{ -i \sum_a \omega^a(x) T^a \right\}.
\end{equation}

A maximum of Eq.~(\ref{eqn:f1}) implies that $\sum_\mu \del_\mu A_\mu
= 0$ up to \oa{2}.  To ensure that gauge dependent quantities are also
\oa{2} improved, we implement the analogous \oa{2} improved gauge fixing
functional
\beq
{\cal F}_{\text{Imp}}^G = \sum_{\mu, x}\frac{1}{2u_0} \mbox{Tr} \, \left\{
	\frac{4}{3} U_\mu^G(x) 
	- \frac{1}{12 u_0}U_\mu^G(x) U_\mu^G(x+\muhat)
 	+ \text{h.c.} \right\}
\eeq
as described in~\cite{LandauGaugeDE}.
We employ a Conjugate Gradient, Fourier Accelerated gauge fixing algorithm
\cite{CandM2} optimally designed for parallel machines.

\section{The Landau Gauge Gluon Propagator}

The gauge links $U_\mu(x)$ are expressed in terms of the continuum
gluon fields as
\beq
U_\mu(x) = {\cal P} e^{ig_0\int_x^{x+\muhat}A_\mu(z)dz} \, ,
\eeq
where ${\cal P}$ denotes path ordering.  From this, the dimensionless
lattice gluon field $A_{\mu}(x)$ may be obtained via
\beq
A_\mu(x+\muhat/2) = \frac{1}{2ig_0}\left(U_\mu(x)-U^{\dagger}_\mu(x)\right)
 - \frac{1}{6ig_0}\Tr\left(U_\mu(x)-U^{\dagger}_\mu(x)\right) \, ,
\label{gluonField}
\eeq
accurate to ${\cal O}(a^2)$.  \oa{2} improved gluon field
operators have also been investigated.  While the infrared behavior is
unaffected by the improvement, the ultraviolet behavior suffers due
to the extended nature of an improved operator.  These results will
be discussed in detail elsewhere \cite{future}.

We calculate the gluon propagator in coordinate space
\beq
D_{\mu\nu}^{ab}(x,y) \equiv \langle \, A_\mu^a(x) \, A_\nu^b(y) \,
\rangle \, , 
\eeq
using (\ref{gluonField}).  To improve statistics, we use translational
invariance and calculate
\beq
D_{\mu\nu}^{ab}(y) = \frac{1}{V} \, \langle \, \sum_x A_\mu^a(x)
A_\nu^b(x + y) \, \rangle \, .
\eeq 
In this report we focus on the scalar part of the propagator,
\beq
D(y) = \frac{1}{N_d-1} \sum_\mu \frac{1}{N_c^2-1} \sum_a D_{\mu\mu}^{aa}(y),
\eeq
where $N_d = 4$ and $N_c = 3$ are the number of dimensions and colors.
This is then Fourier transformed into momentum space 
\beq
D(q) = \sum_y e^{i \qhat \cdot y} D(y) \, ,
\eeq
where the available momentum values, $\qhat$, are given by
\beq
\qhat_\mu  = \frac{2 \pi n_\mu}{a L_\mu}, \qquad
n_\mu \in  \left( -\frac{L_\mu}{2}, \frac{L_\mu}{2} \right] \, ,
\eeq
and $L_\mu$ is the length of the box in the $\mu$ direction.  In the 
continuum, the scalar function $D(q^2)$ is related to the Landau gauge
gluon propagator via
\beq
D_{\mu\nu}^{ab}(q) =
(\delta_{\mu\nu}-\frac{q_{\mu}q_{\nu}}{q^2})\delta^{ab}D(q^2) \, .
\label{eq:landau_prop}
\eeq

The bare, dimensionless lattice gluon propagator, $D(qa)$ is related to the 
renormalized continuum propagator, $D_R(q;\mu)$ via
\beq
a^2 D(qa) = Z_3(\mu,a)D_R(q;\mu),
\eeq
where the renormalization constant, $Z_3(\mu,a)$ is determined by imposing
a renormalization condition at some chosen renormalization scale, $\mu$,
e.g.,
\beq
D_R(q) |_{q^2 = \mu^2} = \frac{1}{\mu^2}. 
\eeq
This means that there is an undetermined multiplicative renormalization 
factor, $Z_3(a)$, on each of our propagators.  Since our purpose is to 
compare our two improved, coarse lattices with the unimproved, finer one,
it is sufficient to consider only the their relative renormalizations.
We have slightly rescaled the improved propagators so as to provide a 
reasonable match with old one.
The relevant quantity is
\beq
Z_3(0.10) / Z_3(0.35) = 1.02.
\eeq

\section{Analysis of Lattice Artifacts}
\label{sec:analysis}

\begin{figure}[p]
\begin{center}
\epsfig{figure=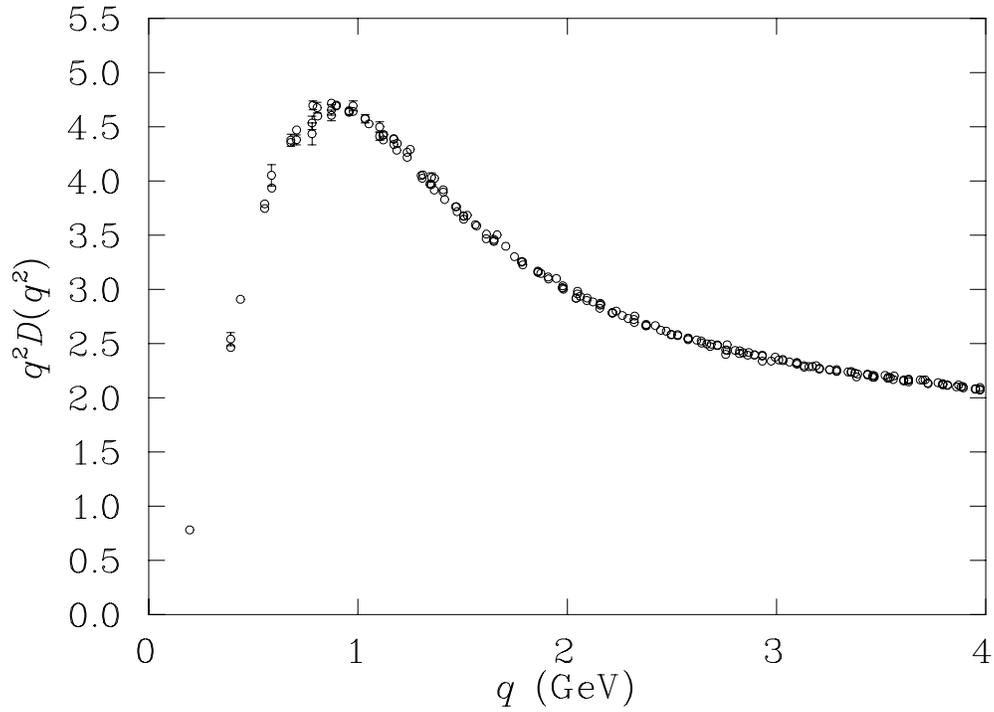,angle=90,width=13cm}
\end{center}
\caption{Gluon propagator from 75 standard, Wilson configurations, on a 
$32^3 \times 64$ lattice with spacing $a=0.10 \text{ fm.}$}
\label{fig:OldPropCut}
\end{figure}

\begin{figure}[p]
\begin{center}
\epsfig{figure=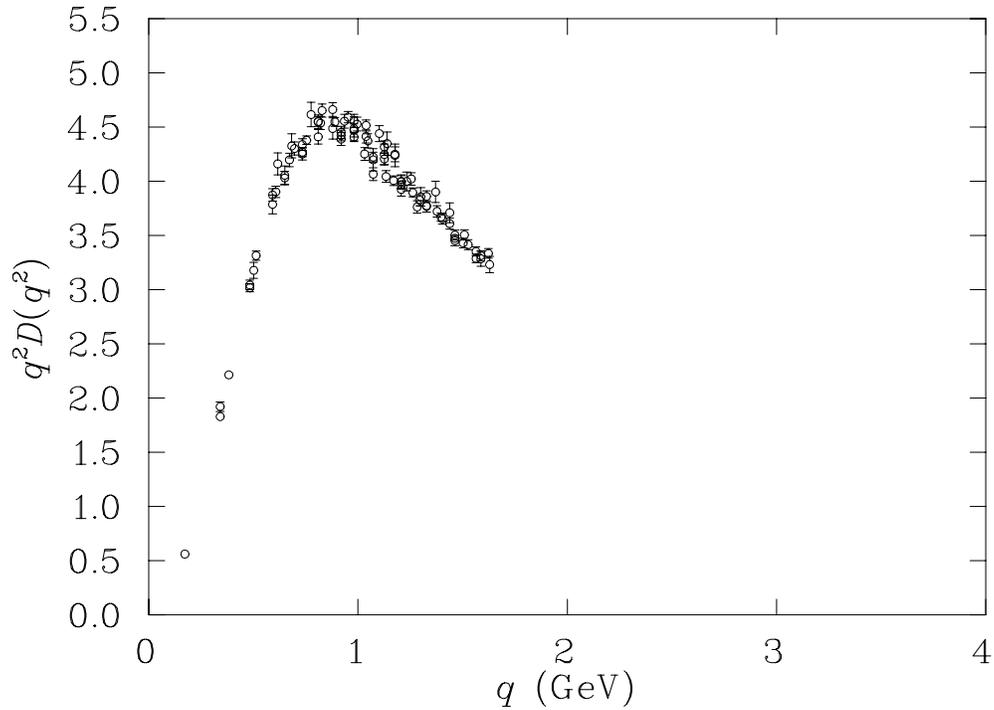,angle=90,width=13cm}
\end{center}
\caption{Gluon propagator from 75 tree-level improved configurations on a
$10^3 \times 20$ lattice with spacing
$a=0.35 \text{ fm,}$ and a physical volume of $3.5^3 \times 7 \text{ fm}^4.$}
\label{fig:ImpSmallCut}
\end{figure}

\begin{figure}[tp]
\begin{center}
\epsfig{figure=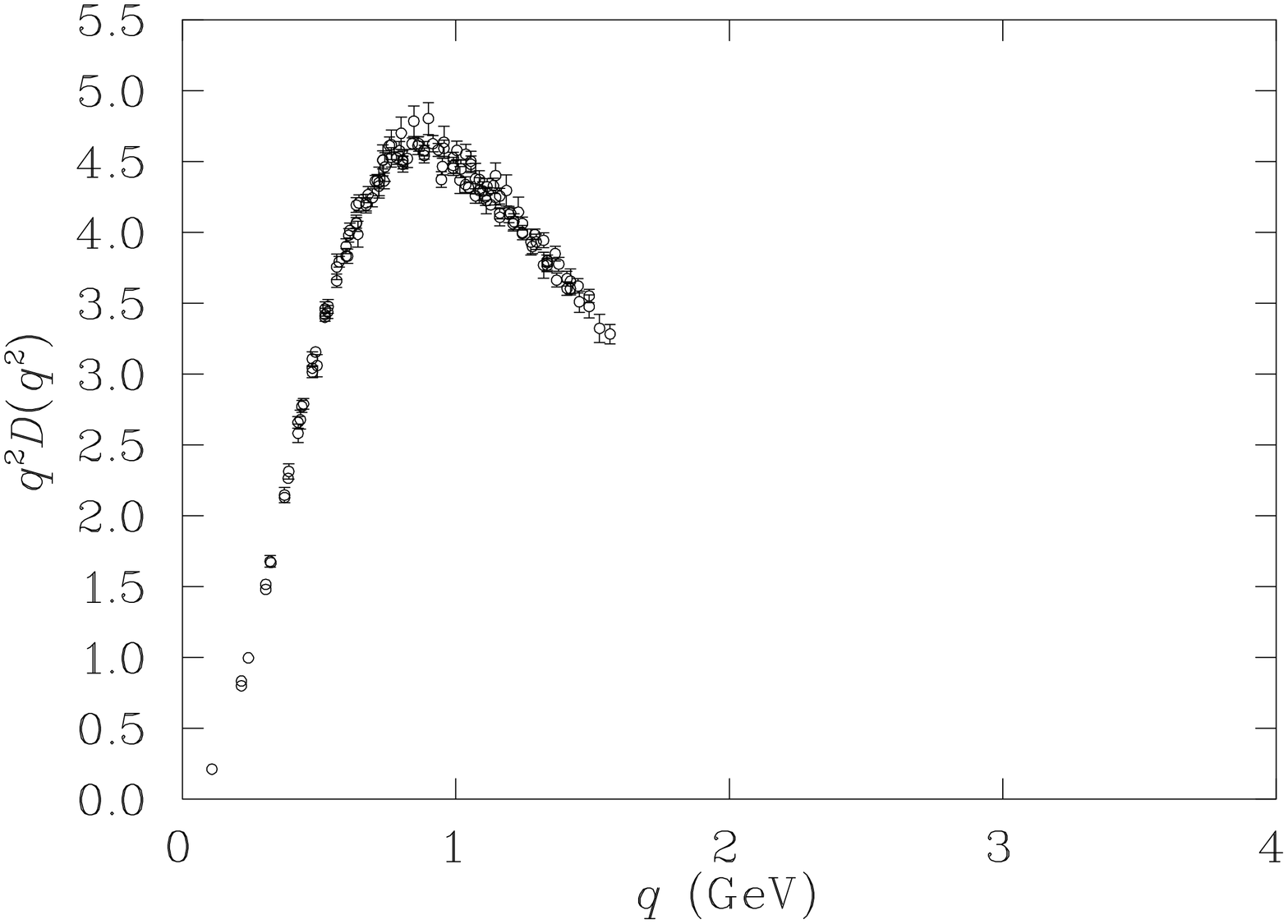,angle=90,width=13cm}
\end{center}
\caption{Gluon propagator from 75 tree-level improved configurations, on a
$16^3 \times 32$ lattice with spacing $a=0.35 \text{ fm,}$ and a physical 
volume of $5.6^3 \times 11.2 \text{ fm}^4.$}
\label{fig:ImpLargeCut}
\end{figure}

To emphasize the nonperturbative behavior of the gluon propagator, we
divide the propagator by the tree level result of lattice perturbation
theory.  Hence Figs.~\ref{fig:OldPropCut}, \ref{fig:ImpSmallCut} and
\ref{fig:ImpLargeCut} are plotted with $q^2 D(q^2)$ on the $y$-axis,
which is expected to approach a constant up to logarithmic corrections
as $q^2\to\infty$.  Here 
\beq
q_{\mu} \equiv \frac{2}{a} \sin\frac{\qhat_{\mu} a}{2}\, .
\label{eq:latt_momenta}
\eeq

To reduce ultraviolet noise resulting from the lattice discretization,
the available momenta are cut half way into the Brillouin zone, that is
\beq
q_{\text{max}} = \frac{\pi}{2a} \, .
\eeq
All figures have a cylinder cut imposed upon them, i.e.\ all momenta
must lie within a cylinder of radius two spatial momentum units
centered about the lattice diagonal.
The propagators are plotted in physical units, which we obtain from
the string tension \cite{zanotti} with $\sqrt{\sigma} = 440$ MeV.

Fig.~\ref{fig:OldPropCut} is reproduced from Ref.~\cite{long_glu},
where the standard, Wilson action is used.  The propagator is
calculated on a $32^3 \times 64$ lattice at $\beta = 6.0$, which
corresponds to a lattice spacing of 0.1 fm.  This propagator produces
the correct asymptotic behavior and a detailed analysis shows the 
anisotropic finite volume errors are small.  However, it was
impossible to rule out isotropic finite volume artifacts.

We use the improved action described above to calculate the gluon
propagator on a small ($10^3 \times 20$), coarse ($a = 0.35$ fm)
lattice, which is shown in Fig.~\ref{fig:ImpSmallCut}.  Despite the
coarse lattice spacing we see that it reproduces the infrared
behavior of Fig.~\ref{fig:OldPropCut}.

Finally, we calculate the propagator on a $16^3 \times 32$ lattice, at
the same $\beta$ providing $a = 0.35$ fm.  This corresponds to a very
large physical volume of $5.6^3 \times 11.2 \text{ fm}^4.$
Fig.~\ref{fig:ImpLargeCut} illustrates the results.  These results
largely agree with the previous two calculations of the propagators.
The behavior of the gluon propagator is not changed by changing the
volume.

\begin{figure}[tp]
\begin{center}
\epsfig{figure=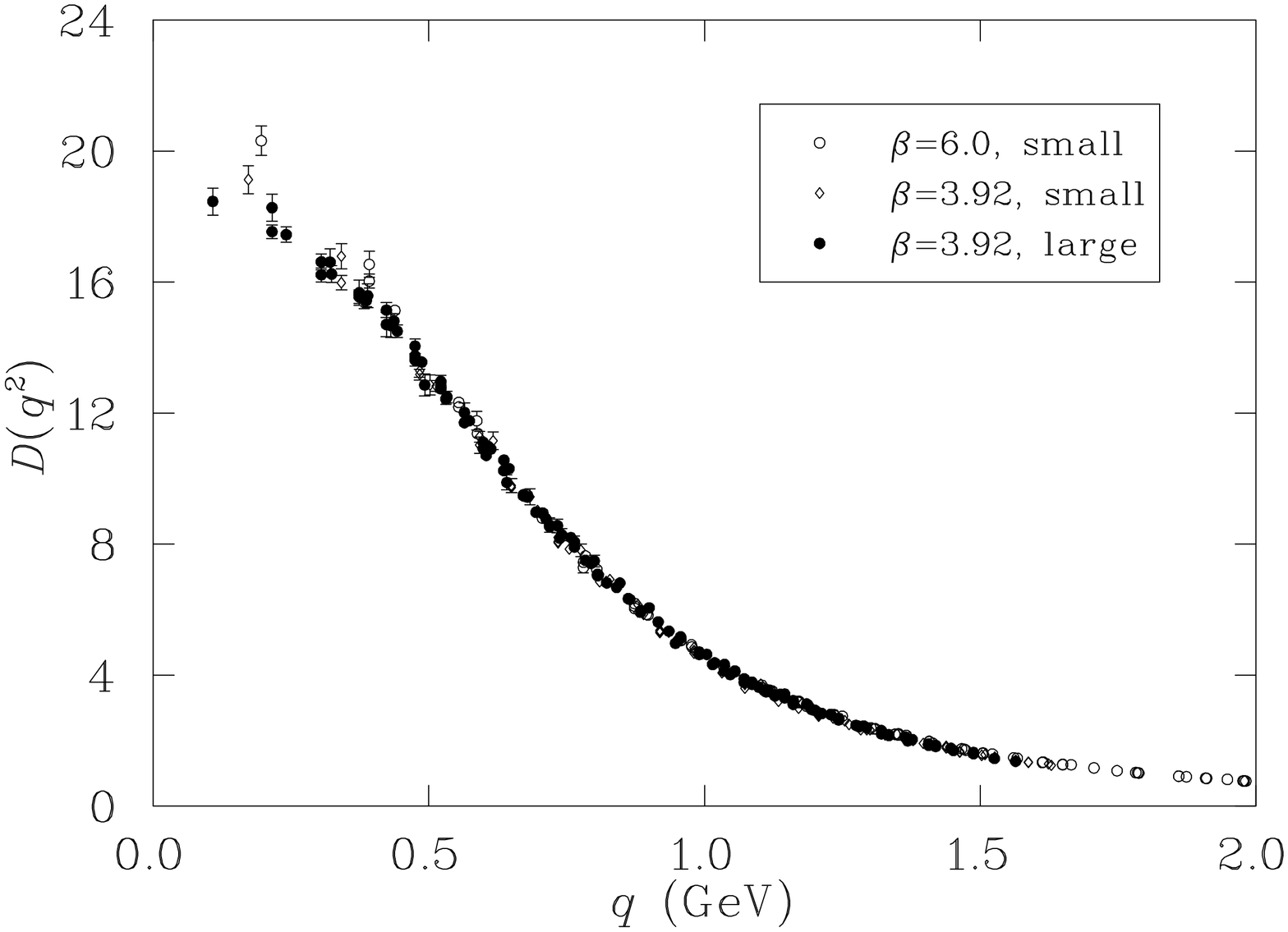,angle=90,width=13cm}
\end{center}
\caption{Comparison of the gluon propagator on the three different lattices.
The volumes are $3.2^3 \times 6.4 \text{ fm}^4,$ $3.5^3 \times 7.0 
\text{ fm}^4,$ and $5.6^3 \times 11.2 \text{ fm}^4.$}
\label{fig:AllProps}
\end{figure}

In Fig.~\ref{fig:AllProps} we have superimposed the gluon propagators
for all three lattices.  Here we plot $D(q^2)$ on the $y$-axis to
allow an alternative examination of the most infrared momenta.  The
points at $\gtrsim$ 350 MeV are very robust with respect to volume.
Only the very lowest momenta points show signs of finite volume
effects.  With more volumes it should be possible to extrapolate to
the infinite volume limit.  We see that the propagator, at the very lowest
momentum points, decreases as the volume increases.  This strongly
suggests an infrared finite propagator.

\section{Conclusions}

The gluon propagator has been calculated on a coarse lattice with an
\oa{2} improved action, in \oa{2} improved Landau gauge.  The infrared
behavior of this propagator is consistent with that of a previous
study on a finer lattice with an unimproved action, but comparable
volume.  The propagator was then calculated on another improved
lattice with the same spacing, but larger volume.  The increase in
volume left the propagator largely unchanged.  In particular, it has
been shown that the turnover observed in \cite{long_glu} is not a
finite volume effect.

With more lattices it may be possible to extrapolate to infinite
volume, but from this study we can only make tentative conclusions.
We have ruled out the $q^{-4}$ behavior popular in Dyson-Schwinger
studies, and any infrared singularity appears to be unlikely.  An
infrared finite propagator is most plausible.  The gluon propagator
would need to drop rapidly for momenta below $\sim$ 350 MeV in order
to vanish as suggested by Zwanziger\cite{Zwan91} and others.  Even
larger volume lattices will be needed to study this possibility.  The
possible effects of lattice Gribov copies remains a very interesting
question and we plan to study this in the near future.

\section*{Acknowledgements}

This work was supported by the Australian Research Council and by
grants of supercomputer time on the CM-5 made available through the
South Australian Centre for Parallel Computing. 
The authors also wish to thank J-I. Skullerud for his useful comments.


\end{document}